\documentclass[a4paper,leqno]{llncs}

\makeatletter

\let\proof\@undefined
\let\endproof\@undefined
\makeatother

\newcommand{\keywords}[1]{\par\addvspace\baselineskip
\noindent\keywordname\enspace\ignorespaces#1}

\usepackage{amsmath}
\usepackage{amssymb}
\usepackage[usenames,dvipsnames]{color}
\usepackage{enumitem}
\usepackage{url}
\usepackage{comment}

\newcommand{\mybox}[1]{\ensuremath{\text{\mbox{\ensuremath{#1}}}}}
\newcommand{\spacedwith}[2]{\ensuremath{\phantom{#2}#1\phantom{#2}}}
\newcommand{\spaced}[1]{\spacedwith{#1}{m}}

\newcommand{\primed}[1]{\ensuremath{#1^{\prime}}}
\newcommand{\dprimed}[1]{\ensuremath{#1^{\prime\prime}}}
\newcommand{\defn}{\spacedwith{\stackrel{\mathbf{def}}{=}}{o}}

\newcommand{\aleq}{\ensuremath{\subseteq}}
\newcommand{\ageq}{\ensuremath{\supseteq}}
\newcommand{\aeq}{\ensuremath{=}}

\newcommand{\askip}{\ensuremath{\textit{skip}}}

\newcommand{\abottom}{\ensuremath{\bot}}
\newcommand{\atopp}{\ensuremath{\top}}
\newcommand{\asemicolon}{\ensuremath{\,;}}

\newcommand{\asemicolontext}{\ensuremath{;}}

\newcommand{\aor}{\ensuremath{\cup}}
\newcommand{\aand}{\ensuremath{\cap}}
\newcommand{\aunion}{\aor}
\newcommand{\aintersect}{\aand}
\newcommand{\alub}{\ensuremath{\bigcup}}
\newcommand{\aglb}{\ensuremath{\bigcap}}
\newcommand{\astar}{\ensuremath{\parallel}}
\newcommand{\kleene}[1]{\ensuremath{{#1}^*}}

\newcommand{\htriple}[3]{\mybox{{#1}\spacedwith{\{#2\}}{\,}{#3}}}
\newcommand{\config}[2]{\mybox{\langle {#1},\,{#2} \rangle}}
\newcommand{\plotkin}[4]{\mybox{\config{#1}{#2}\longrightarrow\config{#3}{#4}}}
\newcommand{\plotkiniter}[5]{\mybox{\config{#1}{#2}\longrightarrow^{#3}\config{#4}{#5}}}
\newcommand{\plotkinstar}[4]{\mybox{\config{#1}{#2}\longrightarrow^*\config{#3}{#4}}}
\newcommand{\milner}[3]{\mybox{{#1}\stackrel{#2}{\longrightarrow}{#3}}}
\newcommand{\kahn}[3]{\mybox{\config{#1}{#2}\longrightarrow{#3}}}
\newcommand{\btriple}[3]{\mybox{{#1}\spacedwith{\ll\mspace{-4mu}{#2}\mspace{-4mu}\gg}{\mspace{-2mu}}{#3}}}
\newcommand{\ftriple}[3]{\mybox{{#1}\spacedwith{[#2]}{\,}{#3}}}
\newcommand{\astriple}[3]{\mybox{{#1}\spacedwith{\##2\#}{\,}{#3}}}
\newcommand{\vtriple}[3]{\mybox{\{{#1}\}\spacedwith{#2}{\,}\{{#3}\}}}
\newcommand{\sembrack}[1]{\ensuremath{[\![#1]\!]}}

\newcommand{\Actions}{\mybox{\mathit{Actions}}}
\newcommand{\Atoms}{\mybox{\mathit{Atoms}}}
\newcommand{\AtomicOperations}{\mybox{\mathit{AtomicOperations}}}
\newcommand{\States}{\mybox{\Sigma}}
\newcommand{\Inconsistent}{\mybox{\mathit{Inconsistent}}}

\newcommand{\s}{\ensuremath{\sigma}}
\newcommand{\sprime}{\primed{\sigma}}
\newcommand{\sdprime}{\dprimed{\s}}
\newcommand{\stprime}{\ensuremath{\s^{\prime\prime\prime}}}

\newcommand{\trace}{\ensuremath{t}}
\newcommand{\tprime}{\primed{t}}

\newcommand{\ic}[1]{\ensuremath{\mathit{ic}(#1)}}
\mathchardef\mhyphen="2D
\newcommand{\icTracesEndingInState}[1]{\ensuremath{\mathit{ic\mhyphen traces\mhyphen ending\mhyphen in\mhyphen state}(#1)}}
\newcommand{\tracesOfView}[1]{\ensuremath{\mathit{traces\mhyphen of\mhyphen view}(#1)}}
\newcommand{\traceSetOfAtomString}[1]{\ensuremath{\mathit{trace\mhyphen set\mhyphen of\mhyphen atom\mhyphen sequence}(#1)}}
\newcommand{\traceSetOfCommand}[1]{\ensuremath{\mathit{trace\mhyphen set\mhyphen of\mhyphen command}(#1)}}

\newcommand{\lfp}[1]{\ensuremath{\mathit{lfp\,}{#1}}}

\newcommand{\Views}{\mybox{\mathit{Views}}}
\newcommand{\vstar}{\ensuremath{*}}
\newcommand{\vunit}{\ensuremath{u}}
\newcommand{\vmodels}{\ensuremath{\models}}
\newcommand{\vlub}{\ensuremath{\bigvee}}
\newcommand{\vglb}{\ensuremath{\bigwedge}}
\newcommand{\vangelic}{\ensuremath{\prec}}
\newcommand{\view}{\ensuremath{v}}
\newcommand{\vprime}{\primed{v}}
\newcommand{\vdprime}{\dprimed{\view}}
\newcommand{\vtprime}{\ensuremath{\view^{\prime\prime\prime}}}
\newcommand{\vone}{\ensuremath{v_1}}
\newcommand{\vtwo}{\ensuremath{v_2}}
\newcommand{\voneprime}{\ensuremath{v_1^\prime}}
\newcommand{\vtwoprime}{\ensuremath{v_2^\prime}}
\newcommand{\erase}[1]{\ensuremath{\lfloor{#1}\rfloor}}
\newcommand{\Axioms}{\mybox{\mathit{Axioms}}}

\newcommand{\sand}{\spaced{\&}}
\newcommand{\siff}{\spaced{\Leftrightarrow}}
\newcommand{\simplies}{\spaced{\Rightarrow}}

\newcommand{\lemmaAlignSpace}{\mbox{}\\[-6.61ex]}

\newcommand{\append}[2]{\ensuremath{{#1}+\!\!\!+\,{#2}}}
\newcommand{\interleave}[2]{\ensuremath{{#1}\otimes{#2}}}

\newcommand{\rulename}[1]{\tag{#1}\label{#1}}

\newcommand{\BEGINMYDEF}{\begin{trivlist}\item}
\newcommand{\ENDMYDEF}{\end{trivlist}}

\newlength{\templength}

\begin{document}

\mainmatter

\title{A Framework for Concurrent Imperative Programming}

\titlerunning{A Framework for Concurrent Imperative Programming}

\author{Stephan van Staden}

\authorrunning{Stephan van Staden}

\institute{ETH Zurich, Switzerland\\
\email{Stephan.vanStaden@inf.ethz.ch}
}

\toctitle{A Framework for Concurrent Imperative Programming}
\tocauthor{Stephan van Staden}

\maketitle

\begin{abstract}
The proposed framework provides a general model of concurrent imperative programming. Programs are modeled as formal languages and concurrency as an interleaving (or shuffle) operator. This yields a simple and elegant algebra of programs. The framework supports the views program logic by Dinsdale-Young and others, which generalizes various type systems and separation logic approaches to program correctness. It also validates familiar operational calculi in small-step and big-step flavours. The consistency of the program logic with respect to the operational rules is established directly and does not use induction on derivations. In fact the whole framework uses only straightforward mathematics. Parametric in states, views and basic commands, it can be instantiated to a variety of concrete languages and settings.
\end{abstract}

\section{Introduction}
A mathematical theory of programming makes it possible to reason about programs and their behaviour in a rigorous way. It abstracts from irrelevant detail, but must still be concrete enough to support all the programming features of interest. For example, modeling programs as binary relations on states (i.e. a program is a set of input-output state pairs) yields a simple and powerful framework for sequential programming. Although it supports reasoning about nondeterministic choice, for example, it abstracts too aggressively to facilitate a compositional treatment of interfering concurrency.

The framework of this paper was inspired by recent work in algebraic semantics and the unification of programming calculi~\cite{wehrman09graphical,hoare09concurrent,hoare12laws,hoare12in,vanstaden12algebra,dinsdale-young12views}. It models programs as (formal) languages over usually infinite alphabets, which makes it possible to model concurrent composition with an interleaving (or shuffle) operator. Languages with interleaving have a simple and elegant algebra that is familiar to many computer scientists. In addition to algebraic reasoning, the paper shows that they also support various deductive and operational calculi of programming.

The framework is parametric in a set of computational states. It provides common programming operators such as sequential composition, nondeterministic choice, iteration and concurrency, which can be combined to model constructs of high-level languages such as if-statements and while-loops. Deductive (i.e. Hoare-style) reasoning comes for free, and requires only a decision about which views~\cite{dinsdale-young12views} to use. Views are abstractions of computational states, and many type systems and separation logic approaches for reasoning about concurrency can be understood in terms of them. By representing states and views as languages, the framework can define the judgement of the views calculus in terms of language operators and relations. It then simple to see why views reasoning works -- the proofs that its rules hold as theorems explain every little detail.

The framework also supports operational calculi, in small-step and big-step flavours, for reasoning about program execution. The small-step calculi are parametric in a set of primitive atomic operations that are easy to implement in a machine. The judgements of the small-step calculi are defined in terms of this set, the language representation of states, and familiar language operators and relations. It is straightforward to prove that the usual operational rules, which show how a computation can proceed in small steps, hold as theorems. The treatment of big-step calculi is similar but simpler, because they describe only the overall result of a computation.

The fact that all judgements have a clear and succinct mathematical meaning makes it straightforward to establish relationships among them. For example, it is possible to show that a deductive calculus is consistent with an operational one without embarking on tedious induction proofs involving derivations with particular sets of rules. The consistency covers any operational rule that is a theorem, and there is no need to revise the proof when new rules, such as optimizations, are adopted. The consistency is also parametric in the chosen computational states, views, and the primitive atomic operations used in execution.

The simplicity of the mathematics will hopefully encourage researchers and practitioners to use the framework for reasoning about high-level languages. The framework could also be useful for investigating and justifying new deductive approaches, program transformations, and operational calculi.

All theorems of this paper have been machine-checked with Isabelle/HOL. A proof script is available online~\cite{onlineproofsAFrameworkForConcurrentImperativeProgramming}.

\paragraph{Outline} Section~\ref{LanguagesAndLaws} summarizes the definition, operators and algebraic laws of languages. Section~\ref{AbstractCalculi} outlines abstract calculi that follow from the algebraic laws. Section~\ref{StatesTraces} covers concepts that are common to the deductive calculi in section~\ref{DeductiveCalculi} and the operational calculi in section~\ref{OperationalCalculi}. The consistency of the deductive and operational calculi follows in section~\ref{Consistency}, and section~\ref{Conclusion} concludes. A treatment of recursion appears as an appendix.

\section{Languages and laws}\label{LanguagesAndLaws}
Formal languages offer a simple and expressive formalism for modeling programs, designs and specifications. They are central to the framework; this section summarizes their basic definitions, operators and algebraic laws. 

An \emph{alphabet} is a set. A \emph{word} over an alphabet $\mathcal{A}$ is a finite sequence of elements from $\mathcal{A}$. A set of words over an alphabet form a \emph{language}. The languages over an alphabet $\mathcal{A}$, being subsets of the set of all words over $\mathcal{A}$, form a complete Boolean algebra. The intended alphabet will always be clear from the context, and is omitted only when the results hold for an arbitrary alphabet. There are several lattice-theoretic constants and operators, for example:
\begin{itemize}
 \item \abottom{} is the empty language $\{\}$.
 \item \atopp{} is the language consisting of all words.
 \item $\alub X$ is the least upper bound of the set of languages $X$, and (\aunion) its binary variant.
 \item $\aglb X$ is the greatest lower bound of languages in $X$, and (\aintersect) its binary variant.
\end{itemize}
Languages also support several other operators that are familiar. In particular:
\begin{itemize}
 \item \askip{} is the language consisting only of the empty word: \\
       $\askip \defn \{[]\}$
 \item (\asemicolontext) is the language concatenation operator: \\
       $P \asemicolon Q \defn \{\append{p}{q} \mid p \in P \sand q \in Q \}$
 \item The Kleene star operator concatenates its argument zero or more times: \\
       $\kleene{P} \defn \alub\{P^n \mid n \in \mathbb{N}\}$ \\
       where \\
       $P^0 \defn \askip$ \\
       $P^{k+1} \defn P \asemicolon P^k$
 \item (\astar) is the language interleaving (or shuffle) operator: \\
       $P \astar Q \defn \alub\{\interleave{p}{q} \mid p \in P \sand q \in Q \}$ \\
       Here \interleave{p}{q} denotes the set of all interleavings of the words $p$ and $q$. It can be defined recursively as follows: \\
       $\interleave{[]}{q} \defn \{q\}$ \\
       $\interleave{p}{[]} \defn \{p\}$ \\
       $\interleave{(e:p)}{(e^\prime:q)} \defn \{e:r \mid r \in \interleave{p}{(e^\prime:q)}\} \cup \{e^\prime:r \mid r \in \interleave{(e:p)}{q}\}$ \\
\end{itemize}

Languages have a rich algebra that simplifies formal reasoning. Some basic properties of the operators appear in Table~\ref{Operators}. 

\begin{table}
\begin{center}
\setlength{\templength}{\tabcolsep}
\setlength{\tabcolsep}{8pt}
\begin{tabular}{l|cccc}
\vrule width 0pt depth 1ex
 & \aor & \aand & \asemicolon & \astar \\
\hline
\vrule width 0pt height 2.5ex
Commutative & yes & yes & no & yes \\
Associative & yes & yes & yes & yes \\
Idempotent & yes & yes & no & no \\
Unit & \abottom & \atopp & \askip & \askip \\
Zero & \atopp & \abottom & \abottom & \abottom
\end{tabular} \\ \textcolor{white}{.}
\caption{Basic properties of the operators.\label{Operators}}
\setlength{\tabcolsep}{\templength}
\end{center}
\end{table}

Since $(\mathbb{P}(\atopp), \aor, \asemicolontext\!, \kleene{}, \abottom, \askip)$ is a Kleene algebra~\cite{kozen94completeness}, all the usual laws and identities hold as theorems. For example, the Kleene star is monotone and satisfies the following laws:
\begin{itemize}
\item $\askip \aor (P \asemicolon \kleene{P})  \aleq  \kleene{P}$
\item	$P \aor (Q \asemicolon R) \aleq R$     \simplies    $\kleene{Q} \asemicolon P \aleq R$
\item	$\askip \aor (\kleene{P} \asemicolon P)  \aleq  \kleene{P}$
\item	$P \aor (R \asemicolon Q) \aleq R$     \simplies    $P \asemicolon \kleene{Q} \aleq R$
\end{itemize}

All the binary operators distribute through (\aunion) and are consequently monotone. A stronger statement is true for $\circ \in \{\aand, \asemicolontext\!, \astar\}$:
\begin{itemize}
\item $P \circ (\alub{}X) \aeq \alub{}\{P \circ Q \mid Q \in X\}$
\item $(\alub{}X) \circ P \aeq \alub{}\{Q \circ P \mid Q \in X\}$
\end{itemize}
The same holds for $\circ = \aunion$ when $X$ is not empty.
In fact it is known that $(\mathbb{P}(\atopp), \aunion, \abottom, \astar, \asemicolontext\!, \askip)$ is a concurrent Kleene algebra~\cite{hoare09concurrent}, and hence (\asemicolontext) and (\astar) interact as follows (the properties also appear in~\cite{bloom96free} as Proposition 5.3 and Corollary 5.4):
\begin{itemize}
 \item $(P \astar Q)\asemicolon(R \astar S) \aleq (P \asemicolon R) \astar (Q \asemicolon S)$
 \item $P \asemicolon (Q \astar R)   \aleq   (P \asemicolon Q) \astar R$
 \item $(P \astar Q) \asemicolon R    \aleq    P \astar (Q \asemicolon R)$
 \item $P \asemicolon Q    \aleq    P \astar Q$
\end{itemize}

In summary, the languages with interleaving satisfy all the laws of programming mentioned in~\cite{hoare12laws,hoare12in,vanstaden12algebra}.

\section{Abstract calculi}\label{AbstractCalculi}
The laws yield abstract versions of several familiar calculi of programming, as discussed in~\cite{hoare09concurrent,hoare12laws,hoare12in,vanstaden12algebra}. The framework will later instantiate selected rules of the abstract calculi to obtain concrete calculi.

\subsection{Hoare logic}
The abstract Hoare triple is defined as follows:
\BEGINMYDEF
$\htriple{P}{Q}{R}  \defn  P \asemicolon Q \aleq R$
\ENDMYDEF
Rules of abstract Hoare logic follow as theorems from the algebra:
\begin{align}
&\rulename{Hskip} \htriple{P}{\askip}{P} \\
&\rulename{Hseq} \htriple{P}{Q}{R}  \sand  \htriple{R}{\primed{Q}}{S}  \simplies  \htriple{P}{Q \asemicolon \primed{Q}}{S} \\
&\rulename{Hchoice} (\forall Q \in X : \htriple{P}{Q}{R})  \simplies  \htriple{P}{\alub X}{R} \\
&\rulename{Hiter} \htriple{P}{Q}{P}  \simplies  \htriple{P}{\kleene{Q}}{P} \\
&\rulename{Hcons} \primed{P} \aleq P \sand \htriple{P}{Q}{R} \sand R \aleq \primed{R} \simplies \htriple{\primed{P}}{Q}{\primed{R}} \\
&\rulename{Hdisj} (\forall P \in X : \htriple{P}{Q}{R}) \simplies \htriple{\alub X}{Q}{R}
\end{align}
The next rules, although interesting, are not used in the sequel.
\begin{align}
&\rulename{Hconj} \htriple{P}{Q}{R}  \sand  \htriple{\primed{P}}{\primed{Q}}{\primed{R}}  \simplies  \htriple{P \aand \primed{P}}{Q \aand \primed{Q}}{R \aand \primed{R}} \\
&\rulename{Hframe} \htriple{P}{Q}{R} \simplies \htriple{F \astar P}{Q}{F \astar R} \\
&\rulename{Hconc} \htriple{P}{Q}{R} \sand \htriple{\primed{P}}{\primed{Q}}{\primed{R}} \simplies \htriple{P \astar \primed{P}}{Q \astar \primed{Q}}{R \astar \primed{R}}
\end{align}

\subsection{Plotkin calculus}
The basic judgement of the abstract small-step calculus of Plotkin is defined in terms of a set of languages called \Actions{}. The only requirement on \Actions{} is that it includes \askip.
\BEGINMYDEF
$\plotkin{P}{s}{P^\prime}{s^\prime} \defn \exists Q \in \Actions : P \ageq Q \asemicolon P^\prime \sand s \asemicolon Q \ageq s^\prime$
\ENDMYDEF
Several rules hold as theorems:
\begin{align}
&\rulename{Paction} P \in \Actions \sand s^\prime \aleq s \asemicolon P \simplies \plotkin{P}{s}{\askip}{s'} \\
&\rulename{Pseq1} \plotkin{\askip \asemicolon P}{s}{P}{s} \\
&\rulename{Pseq2} \plotkin{P}{s}{R}{s^\prime} \simplies \plotkin{P \asemicolon P^\prime}{s}{R \asemicolon P^\prime}{s^\prime} \\
&\rulename{Pchoice} P \in X \simplies \plotkin{\alub X}{s}{P}{s} \\
&\rulename{Piter1} \plotkin{\kleene{P}}{s}{\askip}{s} \\
&\rulename{Piter2} \plotkin{\kleene{P}}{s}{P \asemicolon \kleene{P}}{s} \\
&\rulename{Pconc1} \plotkin{\askip \astar P}{s}{P}{s} \\
&\rulename{Pconc2} \plotkin{P \astar \askip}{s}{P}{s} \\
&\rulename{Pconc3} \plotkin{P}{s}{R}{s^\prime} \simplies \plotkin{P \astar P^\prime}{s}{R \astar P^\prime}{s^\prime} \\
&\rulename{Pconc4} \plotkin{P}{s}{R}{s^\prime} \simplies \plotkin{P^\prime \astar P}{s}{P^\prime \astar R}{s^\prime}
\end{align}

\subsection{Milner calculus}
While the Plotkin calculus hides actions, the abstract Milner calculus makes them explicit:
\BEGINMYDEF
$\milner{P}{Q}{R} \defn Q \in \Actions \sand P \ageq Q \asemicolon R$
\ENDMYDEF
The abstract Milner rules include:
\begin{align}
&\rulename{Maction} P \in \Actions \simplies \milner{P}{P}{\askip} \\
&\rulename{Mseq1} \milner{\askip \asemicolon P}{\askip}{P} \\
&\rulename{Mseq2} \milner{P}{Q}{R} \simplies \milner{P \asemicolon P^\prime}{Q}{R \asemicolon P^\prime} \\
&\rulename{Mchoice} P \in X \simplies \milner{\alub X}{\askip}{P} \\
&\rulename{Miter1} \milner{\kleene{P}}{\askip}{\askip} \\
&\rulename{Miter2} \milner{\kleene{P}}{\askip}{P \asemicolon \kleene{P}} \\
&\rulename{Mconc1} \milner{\askip \astar P}{\askip}{P} \\
&\rulename{Mconc2} \milner{P \astar \askip}{\askip}{P} \\
&\rulename{Mconc3} \milner{P}{Q}{R} \simplies \milner{P \astar P^\prime}{Q}{R \astar P^\prime} \\
&\rulename{Mconc4} \milner{P}{Q}{R} \simplies \milner{P^\prime \astar P}{Q}{P^\prime \astar R}
\end{align}

\subsection{Kahn calculus}
The abstract Kahn calculus is a big-step operational calculus. Consequently, its judgement does not mention actions:
\BEGINMYDEF
$\kahn{P}{s}{s^\prime} \defn s \asemicolon P \ageq s^\prime$
\ENDMYDEF
It is a big-step calculus with several rules:
\begin{align}
&\rulename{Kskip}   \kahn{\askip}{s}{s}   \\
&\rulename{Kseq}   \kahn{P}{s}{s^\prime} \sand \kahn{\primed{P}}{s^\prime}{s^{\prime\prime}}  \simplies  \kahn{P \asemicolon \primed{P}}{s}{s^{\prime\prime}}  \\
&\rulename{Kchoice}  P \in X \sand \kahn{P}{s}{s^\prime} \simplies \kahn{\alub X}{s}{s^\prime} \\
&\rulename{Kiter1}   \kahn{\kleene{P}}{s}{s}    \\
&\rulename{Kiter2}   \kahn{P}{s}{s^\prime} \sand \kahn{\kleene{P}}{s^\prime}{s^{\prime\prime}}  \simplies  \kahn{\kleene{P}}{s}{s^{\prime\prime}} \\
&\rulename{Kconc1}   \kahn{P}{s}{s^\prime} \sand \kahn{\primed{P}}{s^\prime}{s^{\prime\prime}}  \simplies  \kahn{P \astar \primed{P}}{s}{s^{\prime\prime}} \\
&\rulename{Kconc2}   \kahn{P^\prime}{s}{s^\prime} \sand \kahn{P}{s^\prime}{s^{\prime\prime}} \simplies  \kahn{P \astar \primed{P}}{s}{s^{\prime\prime}}
\end{align}

\section{States, traces, descriptions and atoms}\label{StatesTraces}
The framework is parametric in a set of computational states \States. For example, an instantiation of the framework might choose \States{} to be the set of all functions mapping variables into values. Let \s{}, possibly with decorations, denote an element of \States.

The bulk of the framework uses $\States \times \States$ as the language alphabet. That is, the individual elements of a word are pairs of states. Such a word is called a \emph{trace}. A set of traces, i.e. a language over $\States \times \States$, is called a \emph{description}.

A description whose traces all have length one is called an \emph{atom}.  The atoms are isomorphic to the binary relations on states, which makes them suitable for modeling the state-transformation behaviour of the (possibly nondeterministic) primitive operations of a high-level language. Let \Atoms{} be the set of all atoms, ranged over by $a$. The next abbreviations simplify the presentation later on:
\BEGINMYDEF
$a(\s) \defn \{\sprime \mid [(\s,\sprime)] \in a\}$ \\
$a(S) \defn \alub \{a(\s) \mid \s \in S\}$
\ENDMYDEF
\indent A trace \trace{} is internally consistent, written \ic{\trace}, when the states between adjacent pairs are equal:
\BEGINMYDEF
$\ic{[]} \defn True$ \\
$\ic{[(\s, \sprime)]} \defn True$ \\
$\ic{(\s, \sprime) : (\sdprime, \stprime) : \trace}   \defn   \sprime = \sdprime \sand \ic{(\sdprime, \stprime) : \trace}$
\ENDMYDEF
A globally-viewed execution trace will always be internally consistent. From a local (e.g. thread) perspective, a trace might be inconsistent due to interference from the environment. The use of sequences of state pairs in compositional models of concurrency dates back to Park~\cite{park80on}.

The following function gives the set of all internally consistent traces that end in a particular state:
\BEGINMYDEF
$\icTracesEndingInState{\s}  \defn   \{\trace \mid \ic{\trace}   \sand  \exists \tprime, \sprime : \trace = \append{\tprime}{[(\sprime, \s)]}\}$
\ENDMYDEF
It associates each state with a description. Note that for every state \s, the set \icTracesEndingInState{\s} is non-empty, since for example $[(\s,\s)]$ will always be a member.

Let \Inconsistent{} be the set of all traces that are not internally consistent. An inconsistency in a trace cannot be undone later on: $\Inconsistent \asemicolon P \aleq \Inconsistent$.

\section{Deductive calculi}\label{DeductiveCalculi}
The framework simplifies partial-correctness reasoning by supporting the views framework of Dinsdale-Young and others~\cite{dinsdale-young12views}. This section explains concisely how and why views-based reasoning works. After giving the necessary background, it presents three calculi, where each calculus builds on its precursor. The first calculus builds on the abstract Hoare logic, and the development culminates in the full views calculus.
\subsection{Background: the views framework}
A \emph{view} describes a set of computational states, and can be thought of as a special kind of assertion. Views are special because they can be composed with operators that enjoy specific algebraic properties:
\begin{itemize}
\item $(\Views, \vmodels, \vlub, \vglb)$ is a complete lattice.
\item $(\Views, \vstar, \vunit)$ is a commutative monoid.
\item \vstar{} distributes over \vlub: $\view \vstar \vlub V = \vlub\{\view \vstar \vprime \mid \vprime \in V\}$
\item $(\Views, \vangelic)$ is a preorder.
\item \vmodels-closure: $\vmodels \subseteq \vangelic$
\item \vlub-closure: $(\forall \view \in V : \view \vangelic \vprime) \simplies \vlub V \vangelic \vprime$
\item Locality: $\view \vangelic \vprime \simplies \view \vstar \vdprime \vangelic \vprime \vstar \vdprime$
\end{itemize}

An erasure function $\erase{-}$ maps a view to the set of computational states it describes. Erasure must satisfy two properties:
\begin{itemize}
\item Monotone: $\view \vangelic \vprime \simplies \erase{\view} \subseteq \erase{\vprime}$
\item Join-homomorphism: $\erase{\vlub V} = \bigcup\{\erase{\view} \mid \view \in V\}$
\end{itemize}

Finally, the views framework is parametric in a set $\Axioms \subseteq \Views \times \Atoms \times \Views$. The axioms describe how chosen atoms transform views. Each axiom must be sound in the following sense:
\begin{itemize}
\item $(\view, a, \vprime) \in \Axioms \simplies a(\erase{\view \vstar \vdprime}) \subseteq \erase{\vprime \vstar \vdprime}$
\end{itemize}

\subsection{Basic views calculus}
Each view is associated with the set of internally consistent traces that satisfy it at the end, i.e. the traces that establish the view:
\BEGINMYDEF
$\tracesOfView{\view} \defn \alub\{\icTracesEndingInState{\s} \mid \s \in \erase{\view}\}$
\ENDMYDEF
This representation of views as descriptions plays nicely with erasure:
\begin{lemma}\label{TracesAndErasure}
$\tracesOfView{\view} \aleq \tracesOfView{\vprime}    \siff   \erase{v} \subseteq \erase{\vprime}$
\end{lemma}

The definition of the basic views triple uses the abstract Hoare triple and the mapping of views to descriptions:
\BEGINMYDEF
$\btriple{\view}{P}{\vprime} \defn \htriple{(\tracesOfView{\view} \aunion \Inconsistent)}{P}{(\tracesOfView{\vprime} \aunion \Inconsistent)}$
\ENDMYDEF
The judgement \btriple{\view}{P}{\vprime} asserts that, whenever a trace that established \view{} is extended with a trace of $P$ in a consistent way, the resulting trace will establish~\vprime. 

The triple has a simple characterization when $P$ is an atom:
\begin{lemma}\label{BasicTripleForAtom}
$\btriple{\view}{a}{\vprime} \siff a(\erase{\view}) \subseteq \erase{\vprime}$
\end{lemma}
Since the axioms are sound, and \vstar{} has a unit, it is easy to prove:
\begin{align}
&\rulename{Batom} (\view, a, \vprime) \in \Axioms \simplies \btriple{\view}{a}{\vprime}
\end{align}

Several rules of the basic views calculus follow immediately as theorems from their counterparts in the abstract Hoare logic. In particular:
\begin{align}
&\rulename{Bskip} \btriple{\view}{\askip}{\view} \\
&\rulename{Bseq} \btriple{\view}{P}{\vprime} \sand \btriple{\vprime}{P^\prime}{\vdprime} \simplies \btriple{\view}{P \asemicolon P^\prime}{\vdprime} \\
&\rulename{Bchoice} (\forall P \in X : \btriple{\view}{P}{\vprime}) \simplies \btriple{\view}{\alub X}{\vprime}\\
&\rulename{Biter} \btriple{\view}{P}{\view} \simplies \btriple{\view}{\kleene{P}}{\view}
\end{align}

Because erasure is monotone, the rule of consequence follows from Lemma~\ref{TracesAndErasure} and \eqref{Hcons}:
\begin{align}
&\rulename{Bcons} \view \vangelic \vprime \sand \btriple{\vprime}{P}{\vdprime} \sand \vdprime \vangelic \vtprime \simplies \btriple{\view}{P}{\vtprime}
\end{align}

Erasure is a join-homomorphism, which implies:
\begin{lemma}\label{tracesOfViewJoinHomomorphism}
$\tracesOfView{\vlub V} = \bigcup\{\tracesOfView{\view} \mid \view \in V\}$
\end{lemma}
The basic rule of disjunction follows from \eqref{Hcons}, \eqref{Hdisj}, Lemma~\ref{tracesOfViewJoinHomomorphism} and the fact that $\Inconsistent \asemicolon P \aleq \Inconsistent$.
\begin{align}
&\rulename{Bdisj} (\forall \view \in V : \btriple{\view}{P}{\vprime}) \simplies \btriple{\vlub V}{P}{\vprime}
\end{align}

The judgement of the basic calculus is rather weak and easy to establish. This has two important consequences. Firstly, does not constrain the description (the middle operand of the triple) much, so the calculus has broad applicability. Secondly, it does not support reasoning that relies on stronger assumptions. The frame and concurrency rules are therefore missing from the basic calculus. The next two calculi will support more sophisticated reasoning by adopting stronger judgements.

\subsection{Framing calculus}
The framing calculus supports top-level framing. Its judgement uses the judgement of the basic views calculus:
\BEGINMYDEF
$\ftriple{\view}{P}{\vprime} \defn \forall \vdprime \in \Views : \btriple{(\view \vstar \vdprime)}{P}{(\vprime \vstar \vdprime)}$
\ENDMYDEF
The judgement is stronger than the basic one because \vstar{} has a unit.
\begin{theorem}\label{FramingTripleStronger}
$\ftriple{\view}{P}{\vprime} \simplies \btriple{\view}{P}{\vprime}$
\end{theorem}

The rule for atoms follows from the soundness of the axioms and Lemma~\ref{BasicTripleForAtom}:
\begin{align}
&\rulename{Fatom} (\view, a, \vprime) \in \Axioms \simplies \ftriple{\view}{a}{\vprime}
\end{align}

Several rules follow directly from the corresponding ones in the basic calculus:
\begin{align}
&\rulename{Fskip} \ftriple{\view}{\askip}{\view} \\
&\rulename{Fseq} \ftriple{\view}{P}{\vprime} \sand \ftriple{\vprime}{P^\prime}{\vdprime} \simplies \ftriple{\view}{P \asemicolon P^\prime}{\vdprime} \\
&\rulename{Fchoice} (\forall P \in X : \ftriple{\view}{P}{\vprime}) \simplies \ftriple{\view}{\alub X}{\vprime}\\
&\rulename{Fiter} \ftriple{\view}{P}{\view} \simplies \ftriple{\view}{\kleene{P}}{\view}
\end{align}

The rule of consequence holds by Locality and \eqref{Bcons}.
\begin{align}
&\rulename{Fcons} \view \vangelic \vprime \sand \ftriple{\vprime}{P}{\vdprime} \sand \vdprime \vangelic \vtprime \simplies \ftriple{\view}{P}{\vtprime}
\end{align}

The rule \eqref{Bdisj}, together with the fact that \vstar{} is commutative and distributes over \vlub, imply the rule of disjunction:
\begin{align}
&\rulename{Fdisj} (\forall \view \in V : \ftriple{\view}{P}{\vprime}) \simplies \ftriple{\vlub V}{P}{\vprime}
\end{align}

The frame rule follows from the associativity of \vstar.
\begin{align}
&\rulename{Fframe} \ftriple{\view}{P}{\vprime} \simplies \ftriple{\view \vstar \vdprime}{P}{\vprime \vstar \vdprime}
\end{align}

The framing calculus supports only top-level framing -- it does not constrain what happens at the intermediate steps of a computation. Compositional reasoning about concurrency usually demands internal framing, which ensures that concurrent components do not interfere with each other's views. The next calculus gains the concurrency rule by doing exactly this.

\subsection{Full views calculus}
The full views calculus does not reason directly about descriptions. Instead, users of the calculus (e.g. high-level languages) must represent a description as a \emph{command}, which is a language over atoms. Factoring a description into atoms provides a simple way for the calculus to get a handle on its internal structure. As a result, the calculus can offer a compositional rule for concurrency.

The mapping from commands to descriptions is straightforward. Every atom sequence has an associated trace set:
\BEGINMYDEF
$\traceSetOfAtomString{[]}  \defn  \askip$ \\
$\traceSetOfAtomString{a:as}  \defn  a \asemicolon \traceSetOfAtomString{as}$
\ENDMYDEF
Likewise, every command also has a corresponding trace set, which is the description it denotes:
\BEGINMYDEF
$\traceSetOfCommand{C}   \defn    \alub\{\traceSetOfAtomString{as} \mid as \in C\}$
\ENDMYDEF
Let $a$, when used as a command, denote the singleton language $\{[a]\}$.

The judgement of the full views calculus uses an auxiliary judgement for atom sequences:
\BEGINMYDEF
$\astriple{\view}{[]}{\vprime} \defn \ftriple{\view}{\askip}{\vprime}$ \\
$\astriple{\view}{(a:as)}{\vprime} \defn \exists \vdprime \in \Views : \ftriple{\view}{a}{\vdprime} \sand \astriple{\vdprime}{as}{\vprime}$
\ENDMYDEF
The definition of the main judgement quantifies over all the atom sequences of a command:
\BEGINMYDEF
$\vtriple{\view}{C}{\vprime} \defn \forall as \in C : \astriple{\view}{as}{\vprime}$
\ENDMYDEF
\indent The new judgements are stronger than the judgement of the framing calculus in the obvious sense. By induction on $as$, it follows from \eqref{Fseq} that:
\begin{lemma}
$\astriple{\view}{as}{\vprime} \simplies \ftriple{\view}{\traceSetOfAtomString{as}}{\vprime}$
\end{lemma}
This lemma, together with \eqref{Fchoice}, imply:
\begin{theorem}\label{FullTripleStronger}
$\vtriple{\view}{C}{\vprime} \simplies \ftriple{\view}{\traceSetOfCommand{C}}{\vprime}$
\end{theorem}

Most rules of the full views calculus rely on lemmas about the auxiliary judgement. These lemmas are typically proved by induction on atom sequences. Standard mathematical machinery, such as induction, will not be mentioned in the text below. Only the important ingredients of a proof are made explicit, such as direct or indirect dependencies on the properties of views. If a rule immediately follows a lemma, then it is a trivial corollary.

Using \eqref{Fatom} and \eqref{Fskip}, it is simple to establish:
\begin{align}
&\rulename{Vatom} (\view, a, \vprime) \in \Axioms \simplies \vtriple{\view}{a}{\vprime}
\end{align}
The rule \eqref{Fskip} immediately implies:
\begin{align}
&\rulename{Vskip} \vtriple{\view}{\askip}{\view}
\end{align}
By \eqref{Fseq}, it holds that:
\begin{lemma}
$\astriple{\view}{as}{\vprime} \sand \astriple{\vprime}{as^\prime}{\vdprime} \simplies \astriple{\view}{(\append{as}{as^\prime})}{\vdprime}$
\end{lemma}\lemmaAlignSpace
\begin{align}
&\rulename{Vseq} \vtriple{\view}{C}{\vprime} \sand \vtriple{\vprime}{C^\prime}{\vdprime} \simplies \vtriple{\view}{C \asemicolon C^\prime}{\vdprime}
\end{align}
It is trivial to establish the rule for nondeterministic choice:
\begin{align}
&\rulename{Vchoice} (\forall C \in Y : \vtriple{\view}{C}{\vprime}) \simplies \vtriple{\view}{\alub Y}{\vprime}
\end{align}
By \eqref{Vskip} and \eqref{Vseq}, the following lemma holds:
\begin{lemma}
$\vtriple{\view}{C}{\view} \simplies \vtriple{\view}{C^n}{\view}$
\end{lemma}
This lemma, together with \eqref{Vchoice}, imply the rule for iteration:
\begin{align}
&\rulename{Viter} \vtriple{\view}{C}{\view} \simplies \vtriple{\view}{\kleene{C}}{\view}
\end{align}
The rule \eqref{Fcons} and the reflexivity of \vangelic{} can be used to prove:
\begin{lemma}
$\view \vangelic \vprime \sand \astriple{\vprime}{as}{\vdprime} \sand \vdprime \vangelic \vtprime \simplies \astriple{\view}{as}{\vtprime}$
\end{lemma}\lemmaAlignSpace
\begin{align}
&\rulename{Vcons} \view \vangelic \vprime \sand \vtriple{\vprime}{C}{\vdprime} \sand \vdprime \vangelic \vtprime \simplies \vtriple{\view}{C}{\vtprime}
\end{align}
Using \eqref{Fdisj}, the fact that $(\Views, \vmodels, \vlub, \vglb)$ is a complete lattice, \vmodels-closure, the reflexivity of \vangelic{}, and \eqref{Fcons}, it is possible to establish:
\begin{lemma}
$(\forall \view \in V : \astriple{\view}{as}{\vprime}) \simplies \astriple{\vlub V}{as}{\vprime}$
\end{lemma}\lemmaAlignSpace
\begin{align}
&\rulename{Vdisj} (\forall \view \in V : \vtriple{\view}{C}{\vprime}) \simplies \vtriple{\vlub V}{C}{\vprime}
\end{align}
From \eqref{Fframe} follows:
\begin{lemma}
$\astriple{\view}{as}{\vprime} \simplies \astriple{(\view \vstar \vdprime)}{as}{(\vprime \vstar \vdprime)}$
\end{lemma}\lemmaAlignSpace
\begin{align}
&\rulename{Vframe} \vtriple{\view}{C}{\vprime} \simplies \vtriple{\view \vstar \vdprime}{C}{\vprime \vstar \vdprime}
\end{align}
Using \eqref{Fframe}, \eqref{Fseq} and the commutativity of \vstar, one can show:
\begin{lemma}
$\astriple{\vone}{as_1}{\voneprime} \sand \astriple{\vtwo}{as_2}{\vtwoprime} \sand as \in \interleave{as_1}{as_2} \simplies \astriple{(\vone \vstar \vtwo)}{as}{(\voneprime \vstar \vtwoprime)}$
\end{lemma}
This directly yields a compositional rule for concurrency:
\begin{align}
&\rulename{Vconc} \vtriple{\vone}{C_1}{\voneprime} \sand \vtriple{\vtwo}{C_2}{\vtwoprime} \simplies \vtriple{\vone \vstar \vtwo}{C_1 \astar C_2}{\voneprime \vstar \vtwoprime}
\end{align}


\section{Operational calculi}\label{OperationalCalculi}
Program execution can be investigated formally with operational calculi, which help to discover valid executions of programs.

Small-step calculi, such as the Plotkin~\cite{plotkin81structural} and Milner~\cite{milner80calculus} ones, are concerned with how a computation can unfold by performing (a sequence of) actions that are easy to implement in a computer. These calculi are therefore parametric in a set $\AtomicOperations \subseteq \Atoms$, whose elements model small atomic steps. For example, a hypothetical high-level language might include Boolean tests, variable assignments and heap operations in this set. The framework defines \Actions{} in terms of \AtomicOperations:
\BEGINMYDEF
$\Actions \defn \{\askip\} \cup \AtomicOperations$
\ENDMYDEF
Think about \askip{} is the trivial action that takes zero time to execute. It cannot change the computational state. In contrast to this, an action from \AtomicOperations{} embodies real work and may transform the state. Think about its execution as taking, say, one time unit.

Big-step calculi, such as the Kahn's natural semantics~\cite{kahn87natural} (see~\cite{nielson99semantics} for its application to imperative programming), will usually include rules for executing selected actions. However, the emphasis is on the ultimate result of a computation, and not on intermediate steps.

This section develops operational calculi for descriptions and also for commands.

\subsection{Descriptions}
The abstract Plotkin and Kahn calculi do not mention computational states explicitly. However, they can be instantiated to obtain the familiar versions.
\subsubsection{Plotkin calculus}
The judgement of the Plotkin calculus is defined in terms of the abstract one:
\BEGINMYDEF
$\plotkin{P}{\s}{P^\prime}{\sprime} \defn$ \\$\exists \trace \in \icTracesEndingInState{\s}, \tprime \in \icTracesEndingInState{\sprime} : \plotkin{P}{\{\trace\}}{P^\prime}{\{\tprime\}}$
\ENDMYDEF
It says that one way of executing $P$ is to execute some action followed by $P^\prime$. The action itself is hidden -- only its effect on the state is explicit in the judgement.

There is also an equivalent characterization, which expresses that what happened before the initial state does not matter:
\begin{lemma}
$\plotkin{P}{\s}{P^\prime}{\sprime} \siff$ \\$\forall \trace \in \icTracesEndingInState{\s} : \exists \tprime \in \icTracesEndingInState{\sprime} : \plotkin{P}{\{\trace\}}{P^\prime}{\{\tprime\}}$
\end{lemma}
The rules of the Plotkin calculus are all easy to derive. A rule for atomic operations follows from \eqref{Paction}:
\begin{align}
&\rulename{PDatom} a \in \AtomicOperations \sand \sprime \in a(\s) \simplies \plotkin{a}{\s}{\askip}{\sprime}
\end{align}
Other rules are trivial consequences of their abstract counterparts.
\begin{align}
&\rulename{PDseq1} \plotkin{\askip \asemicolon P}{\s}{P}{\s} \\
&\rulename{PDseq2} \plotkin{P}{\s}{R}{\sprime} \simplies \plotkin{P \asemicolon P^\prime}{\s}{R \asemicolon P^\prime}{\sprime} \\
&\rulename{PDchoice} P \in X \simplies \plotkin{\alub X}{\s}{P}{\s} \\
&\rulename{PDiter1} \plotkin{\kleene{P}}{\s}{\askip}{\s} \\
&\rulename{PDiter2} \plotkin{\kleene{P}}{\s}{P \asemicolon \kleene{P}}{\s} \\
&\rulename{PDconc1} \plotkin{\askip \astar P}{\s}{P}{\s} \\
&\rulename{PDconc2} \plotkin{P \astar \askip}{\s}{P}{\s} \\
&\rulename{PDconc3} \plotkin{P}{\s}{R}{\sprime} \simplies \plotkin{P \astar P^\prime}{\s}{R \astar P^\prime}{\sprime} \\
&\rulename{PDconc4} \plotkin{P}{\s}{R}{\sprime} \simplies \plotkin{P^\prime \astar P}{\s}{P^\prime \astar R}{\sprime}
\end{align}
The definitions of the iterated and reflexive transitive versions of the judgement are standard:
\BEGINMYDEF
$\plotkiniter{P}{\s}{0}{P^\prime}{\sprime} \defn P \aeq P^\prime \sand \s = \sprime$ \\
$\plotkiniter{P}{\s}{n+1}{P^\prime}{\sprime} \defn \exists P^{\prime\prime}, \sdprime : \plotkin{P}{\s}{P^{\prime\prime}}{\sdprime} \sand \plotkiniter{P^{\prime\prime}}{\sdprime}{n}{P^\prime}{\sprime}$ \\
$\plotkinstar{P}{\s}{P^\prime}{\sprime} \defn \exists n : \plotkiniter{P}{\s}{n}{P^\prime}{\sprime}$
\ENDMYDEF

\subsubsection{Kahn calculus}
The judgement of the Kahn calculus is defined in terms of the abstract Kahn judgement:
\BEGINMYDEF
$\kahn{P}{\s}{\sprime} \defn$\\ $\exists \trace \in \icTracesEndingInState{\s}, \tprime \in \icTracesEndingInState{\sprime} : \kahn{P}{\{\trace\}}{\{\tprime\}}$
\ENDMYDEF
It says that $P$ has an internally consistent trace that can transform the initial state \s{} into the final state \sprime. What brought about the initial state is again unimportant:
\begin{lemma}\label{KahnLemma}
$\kahn{P}{\s}{\sprime} \siff \forall \trace \in \icTracesEndingInState{\s} : \exists \tprime \in \icTracesEndingInState{\sprime} : \kahn{P}{\{\trace\}}{\{\tprime\}}$
\end{lemma}
It is straightforward to obtain a rule for executing atomic operations:
\begin{align}
&\rulename{KDatom} a \in \AtomicOperations \sand \sprime \in a(\s) \simplies \kahn{a}{\s}{\sprime}
\end{align}
Other rules follow trivially from Lemma~\ref{KahnLemma} and their abstract counterparts:
\begin{align}
&\rulename{KDskip}   \kahn{\askip}{\s}{\s}   \\
&\rulename{KDseq}   \kahn{P}{\s}{\sprime} \sand \kahn{\primed{P}}{\sprime}{\sdprime}  \simplies  \kahn{P \asemicolon \primed{P}}{\s}{\sdprime}  \\
&\rulename{KDchoice}  P \in X \sand \kahn{P}{\s}{\sprime} \simplies \kahn{\alub X}{\s}{\sprime} \\
&\rulename{KDiter1}   \kahn{\kleene{P}}{\s}{\s}    \\
&\rulename{KDiter2}   \kahn{P}{\s}{\sprime} \sand \kahn{\kleene{P}}{\sprime}{\sdprime}  \simplies  \kahn{\kleene{P}}{\s}{\sdprime} \\
&\rulename{KDconc1}   \kahn{P}{\s}{\sprime} \sand \kahn{\primed{P}}{\sprime}{\sdprime}  \simplies  \kahn{P \astar \primed{P}}{\s}{\sdprime} \\
&\rulename{KDconc2}   \kahn{P^\prime}{\s}{\sprime} \sand \kahn{P}{\sprime}{\sdprime} \simplies  \kahn{P \astar \primed{P}}{\s}{\sdprime}
\end{align}

\subsubsection{Relationships}
The Plotkin judgement can be characterized in terms of the Milner and Kahn judgements:
\begin{lemma}\label{PlotkinCharacterization}
$\plotkin{P}{\s}{P^\prime}{\sprime} \siff \exists Q : \milner{P}{Q}{P^\prime} \sand \kahn{Q}{\s}{\sprime}$
\end{lemma}
This lemma makes it clear that the Plotkin judgement hides the action that effected the state change.

An equivalent and perhaps more familiar formulation uses a function \sembrack{-} that captures the state-transformation behaviour of a description:
\BEGINMYDEF
$\sembrack{P}(\s)   \defn   \{\sprime \mid \kahn{P}{\s}{\sprime}\}$
\ENDMYDEF
Thus $\sprime \in \sembrack{P}(\s)   \siff   \kahn{P}{\s}{\sprime}$, and the relationship of Lemma~\ref{PlotkinCharacterization} can be written as follows:
\BEGINMYDEF
$\plotkin{P}{\s}{P^\prime}{\sprime} \siff \exists Q : \milner{P}{Q}{P^\prime} \sand \sprime \in \sembrack{Q}(\s)$
\ENDMYDEF
Note that actions have simple state-transformation behaviour: $\sembrack{\askip}(\s) = \{\s\}$ and $\sembrack{a}(\s) = a(\s)$.

Another relationship involves the Plotkin and Kahn judgements. The judgement \plotkinstar{P}{\s}{\askip}{\sprime} says that $P$ can transform the initial state \s{} into the final state \sprime:
\begin{lemma}\label{PlotkinStarApproximatesDescriptions}
$\plotkinstar{P}{\s}{\askip}{\sprime} \simplies \kahn{P}{\s}{\sprime}$
\end{lemma}
It also says that the input/output state transformations described by the reflexive transitive closure of the Plotkin judgement only approximate those of the Kahn judgement. However, this does not mean that the Plotkin judgement is useless. It yields a calculus with interesting rules for concurrency, while the Kahn calculus has only trivial ones. Furthermore, the Plotkin calculus contains information about non-terminating behaviours that the Kahn calculus cannot describe.

\subsection{Commands}
Although a command is only a dressed-up description, it may be desirable to reason directly about its execution. Fortunately, the mapping from commands to descriptions has nice algebraic properties:
\begin{lemma}[Homomorphism]\label{Homomorphism}\mbox{}
\begin{itemize}
 \item $\traceSetOfCommand{a} \aeq a$
 \item $\traceSetOfCommand{\askip} \aeq \askip$
 \item $\traceSetOfCommand{C \asemicolon C^\prime} \aeq \traceSetOfCommand{C} \asemicolon \traceSetOfCommand{C^\prime}$
 \item $\traceSetOfCommand{\alub Y} \aeq \alub\{\traceSetOfCommand{C} \mid C \in Y\}$
 \item $\traceSetOfCommand{\kleene{C}} \aeq \kleene{\traceSetOfCommand{C}}$
 \item $\traceSetOfCommand{C \astar C^\prime} \aeq \traceSetOfCommand{C} \astar \traceSetOfCommand{C^\prime}$
\end{itemize}
\end{lemma}
This will make it easy to construct operational calculi for commands.

\subsubsection{Plotkin calculus}
The Plotkin judgement for commands is defined in terms of the one for descriptions:
\BEGINMYDEF
$\plotkin{C}{\s}{C^\prime}{\sprime} \defn \plotkin{\traceSetOfCommand{C}}{\s}{\traceSetOfCommand{C^\prime}}{\sprime}$
\ENDMYDEF
Several rules for commands follow from the corresponding rules for descriptions and Lemma~\ref{Homomorphism}:
\begin{align}
&\rulename{PCatom} a \in \AtomicOperations \sand \sprime \in a(\s) \simplies \plotkin{a}{\s}{\askip}{\sprime} \\
&\rulename{PCseq1} \plotkin{\askip \asemicolon C}{\s}{C}{\s} \\
&\rulename{PCseq2} \plotkin{C}{\s}{C^\prime}{\sprime} \simplies \plotkin{C \asemicolon C^{\prime\prime}}{\s}{C^\prime \asemicolon C^{\prime\prime}}{\sprime} \\
&\rulename{PCchoice} C \in Y \simplies \plotkin{\alub Y}{\s}{C}{\s} \\
&\rulename{PCiter1} \plotkin{\kleene{C}}{\s}{\askip}{\s} \\
&\rulename{PCiter2} \plotkin{\kleene{C}}{\s}{C \asemicolon \kleene{C}}{\s} \\
&\rulename{PCconc1} \plotkin{\askip \astar C}{\s}{C}{\s} \\
&\rulename{PCconc2} \plotkin{C \astar \askip}{\s}{C}{\s} \\
&\rulename{PCconc3} \plotkin{C}{\s}{C^\prime}{\sprime} \simplies \plotkin{C \astar C^{\prime\prime}}{\s}{C^\prime \astar C^{\prime\prime}}{\sprime} \\
&\rulename{PCconc4} \plotkin{C}{\s}{C^\prime}{\sprime} \simplies \plotkin{C^{\prime\prime} \astar C}{\s}{C^{\prime\prime} \astar C^\prime}{\sprime}
\end{align}
There are also iterated and reflexive transitive versions of the judgement:
\BEGINMYDEF
$\plotkiniter{C}{\s}{0}{C^\prime}{\sprime} \defn C \aeq C^\prime \sand \s = \sprime$ \\
$\plotkiniter{C}{\s}{n+1}{C^\prime}{\sprime} \defn \exists C^{\prime\prime}, \sdprime : \plotkin{C}{\s}{C^{\prime\prime}}{\sdprime} \sand \plotkiniter{C^{\prime\prime}}{\sdprime}{n}{C^\prime}{\sprime}$ \\
$\plotkinstar{C}{\s}{C^\prime}{\sprime} \defn \exists n : \plotkiniter{C}{\s}{n}{C^\prime}{\sprime}$
\ENDMYDEF
and the following relationship holds:
\begin{lemma}\label{PlotkinStarForCommands}
$\plotkinstar{C}{\s}{C^\prime}{\sprime} \simplies$\\ $\plotkinstar{\traceSetOfCommand{C}}{\s}{\traceSetOfCommand{C^\prime}}{\sprime}$
\end{lemma}

\subsubsection{Milner calculus}
The Milner judgement for commands uses the Milner judgement for descriptions:
\BEGINMYDEF
$\milner{C}{C^\prime}{C^{\prime\prime}} \defn \milner{\traceSetOfCommand{C}}{\traceSetOfCommand{C^\prime}}{\traceSetOfCommand{C^{\prime\prime}}}$
\ENDMYDEF
The previous Milner rules and Lemma~\ref{Homomorphism} directly yield rules for commands:
\begin{align}
&\rulename{MCatom} a \in \AtomicOperations \simplies \milner{a}{a}{\askip} \\
&\rulename{MCseq1} \milner{\askip \asemicolon C}{\askip}{C} \\
&\rulename{MCseq2} \milner{C}{C^\prime}{C^{\prime\prime}} \simplies \milner{C \asemicolon C_1}{C^\prime}{C^{\prime\prime} \asemicolon C_1} \\
&\rulename{MCchoice} C \in Y \simplies \milner{\alub Y}{\askip}{C} \\
&\rulename{MCiter1} \milner{\kleene{C}}{\askip}{\askip} \\
&\rulename{MCiter2} \milner{\kleene{C}}{\askip}{C \asemicolon \kleene{C}} \\
&\rulename{MCconc1} \milner{\askip \astar C}{\askip}{C} \\
&\rulename{MCconc2} \milner{C \astar \askip}{\askip}{C} \\
&\rulename{MCconc3} \milner{C}{C^\prime}{C^{\prime\prime}} \simplies \milner{C \astar C_1}{C^\prime}{C^{\prime\prime} \astar C_1} \\
&\rulename{MCconc4} \milner{C}{C^\prime}{C^{\prime\prime}} \simplies \milner{C_1 \astar C}{C^\prime}{C_1 \astar C^{\prime\prime}}
\end{align}

\subsubsection{Kahn calculus}
Here is the Kahn judgement for commands:
\BEGINMYDEF
$\kahn{C}{\s}{\sprime} \defn \kahn{\traceSetOfCommand{C}}{\s}{\sprime}$
\ENDMYDEF
As expected, Lemma~\ref{Homomorphism} is useful for deriving rules from the ones for descriptions:
\begin{align}
&\rulename{KCatom} a \in \AtomicOperations \sand \sprime \in a(\s) \simplies \kahn{a}{\s}{\sprime} \\
&\rulename{KCskip}   \kahn{\askip}{\s}{\s}   \\
&\rulename{KCseq}   \kahn{C}{\s}{\sprime} \sand \kahn{\primed{C}}{\sprime}{\sdprime}  \simplies  \kahn{C \asemicolon \primed{C}}{\s}{\sdprime}  \\
&\rulename{KCchoice}  C \in Y \sand \kahn{C}{\s}{\sprime} \simplies \kahn{\alub Y}{\s}{\sprime} \\
&\rulename{KCiter1}   \kahn{\kleene{C}}{\s}{\s}    \\
&\rulename{KCiter2}   \kahn{C}{\s}{\sprime} \sand \kahn{\kleene{C}}{\sprime}{\sdprime}  \simplies  \kahn{\kleene{C}}{\s}{\sdprime} \\
&\rulename{KCconc1}   \kahn{C}{\s}{\sprime} \sand \kahn{\primed{C}}{\sprime}{\sdprime}  \simplies  \kahn{C \astar \primed{C}}{\s}{\sdprime} \\
&\rulename{KCconc2}   \kahn{C^\prime}{\s}{\sprime} \sand \kahn{C}{\sprime}{\sdprime} \simplies  \kahn{C \astar \primed{C}}{\s}{\sdprime}
\end{align}

\subsubsection{Relationships}
The Plotkin, Milner and Kahn judgements for commands enjoy a similar relationship as before:
\begin{lemma}
$\plotkin{C}{\s}{C^\prime}{\sprime} \siff \exists C^{\prime\prime} : \milner{C}{C^{\prime\prime}}{C^\prime} \sand \kahn{C^{\prime\prime}}{\s}{\sprime}$
\end{lemma}
This can also be formulated in terms of a state-transformation function for commands:
\BEGINMYDEF
$\sembrack{C}(\s) \defn \{\sprime \mid \kahn{C}{\s}{\sprime}\}$
\ENDMYDEF
Note that $\sembrack{C} = \sembrack{\traceSetOfCommand{C}}$, so Lemma~\ref{Homomorphism} can help to characterize the state-transformation behaviour of commands as e.g. equations.

A familiar relationship holds between the Plotkin and Kahn judgements:
\begin{lemma}\label{PlotkinStarApproximatesCommands}
$\plotkinstar{C}{\s}{\askip}{\sprime} \simplies \kahn{C}{\s}{\sprime}$
\end{lemma}
It is a direct consequence of Lemmas~\ref{PlotkinStarForCommands}, \ref{Homomorphism} and \ref{PlotkinStarApproximatesDescriptions}.

\section{Consistency of deductive and operational calculi}\label{Consistency}
The deductive calculi give the standard partial correctness guarantees, and are in this sense consistent with respect to the operational calculi. To establish this formally, it helps to start with the weakest deductive and operational judgements. The proof of the consistency of the basic views calculus and the Kahn calculus is straightforward.
\begin{theorem}
$\btriple{\view}{P}{\vprime} \simplies (\forall \s \in \erase{\view} : \kahn{P}{\s}{\sprime} \simplies \sprime \in \erase{\vprime})$
\end{theorem}
Together with earlier definitions and results, this theorem simplifies the proofs of other consistency statements. For example, the following statement is an immediate consequence:
\begin{corollary}
$\btriple{\view}{P}{\vprime} \simplies (\forall \s \in \erase{\view} : \sembrack{P}(\s)  \subseteq \erase{\vprime})$
\end{corollary}
Lemma~\ref{PlotkinStarApproximatesDescriptions} renders the consistency of the basic views calculus and the Plotkin calculus trivial:
\begin{corollary}
$\btriple{\view}{P}{\vprime} \simplies (\forall \s \in \erase{\view} : \plotkinstar{P}{\s}{\askip}{\sprime} \simplies \sprime \in \erase{\vprime})$
\end{corollary}

The consistency of the framing calculus follows by Theorem~\ref{FramingTripleStronger}.
\begin{corollary}
$\ftriple{\view}{P}{\vprime} \simplies (\forall \s \in \erase{\view} : \kahn{P}{\s}{\sprime} \simplies \sprime \in \erase{\vprime})$
\end{corollary}
\begin{corollary}
$\ftriple{\view}{P}{\vprime} \simplies (\forall \s \in \erase{\view} : \sembrack{P}(\s)  \subseteq \erase{\vprime})$
\end{corollary}
\begin{corollary}
$\ftriple{\view}{P}{\vprime} \simplies (\forall \s \in \erase{\view} : \plotkinstar{P}{\s}{\askip}{\sprime} \simplies \sprime \in \erase{\vprime})$
\end{corollary}

The full views calculus is also consistent with respect to the operational calculi. It is a consequence of Theorem~\ref{FullTripleStronger} and Lemma~\ref{PlotkinStarApproximatesCommands}.
\begin{corollary}
$\vtriple{\view}{C}{\vprime} \simplies (\forall \s \in \erase{\view} : \kahn{C}{\s}{\sprime} \simplies \sprime \in \erase{\vprime})$
\end{corollary}
\begin{corollary}
$\vtriple{\view}{C}{\vprime} \simplies (\forall \s \in \erase{\view} : \sembrack{C}(\s)  \subseteq \erase{\vprime})$
\end{corollary}
\begin{corollary}
$\vtriple{\view}{C}{\vprime} \simplies (\forall \s \in \erase{\view} : \plotkinstar{C}{\s}{\askip}{\sprime} \simplies \sprime \in \erase{\vprime})$
\end{corollary}

\section{Conclusion}\label{Conclusion}
The framework is exactly what its name suggests -- a scaffolding for modeling high-level languages and reasoning about them. Users can choose the state space, views, and the primitive atomic operations. The framework provides common programming operators, such as concurrent and sequential composition, nondeterministic choice and iteration. These operators obey a rich set of algebraic laws, and can be combined to model high-level language constructs such as conditionals and while-loops. The framework provides deductive and operational reasoning for free.

The deductive and operational calculi are largely decoupled, yet still consistent with each other. This consistency is robust with respect to the exact operational rules. For example, it is perfectly acceptable to add the rule:
\begin{align}
&\rulename{PDfuturechoice} \plotkin{P \asemicolon (P^\prime \aor P^{\prime\prime})}{\s}{P \asemicolon P^\prime}{\s}
\end{align}
As long as a new rule is a theorem, it cannot contradict the deductive calculi. The consistency is parametric in \AtomicOperations, so it does not even depend on the exact choice of the primitive machine-executable operations.

The fact that everything is simple mathematics will hopefully encourage researchers and practitioners to propose extensions to the framework. It appears likely that other deductive reasoning techniques for concurrency will be developed in the future, and fitting them into this framework might clearly show why they work. Just like views-based reasoning, a method might require a new judgement, proofs that its rules are theorems, and a demonstration of its consistency with respect to the operational calculi. The framework also makes it possible to justify new program optimizations and calculi for program execution.

\paragraph{Acknowledgements} Special thanks to Tony Hoare for encouragement and fruitful discussions. This work was supported by ETH Research Grant ETH-15 10-1.

\bibliography{Framework}

\section*{Appendix: Recursion}
Recursion is fundamental to procedure and loop mechanisms of high-level languages. The framework allows a standard treatment of recursion, on the level of descriptions and also on the level of commands, in terms of monotone functions and least fixpoints. It validates the usual deductive and operational rules for recursion, which is the main topic of this section.

The Knaster-Tarski theorem says that every monotone function $f$ on a complete lattice has a least fixpoint \lfp{f} (also written $\mu x . f(x)$), and that:
\BEGINMYDEF
$\lfp{f} = \aglb\{P \mid f(P) \aleq P\}$
\ENDMYDEF

\subsection*{Deductive rules}
Suppose a deductive calculus has the following properties:
\begin{enumerate}
 \item The programs (e.g. descriptions or commands) form a complete lattice.
 \item A choice rule holds: showing that all elements of an arbitrary set satisfy a pre/post specification is sufficient for the least upper bound of the set to satisfy it.
 \item Triples are downward-closed in the middle (i.e. program) argument.
\end{enumerate}
Then the weakest satisfiers of pre/post specifications (i.e. specification statements) exist, and the usual rule for recursion is readily proved with the Knaster-Tarski theorem. (This technique is also applied elsewhere, e.g. in~\cite{hoare11locality}.)

The abstract Hoare logic and all three deductive calculi for views have these properties. So the following rules, where $f$ in each of them is a monotone function of an appropriate type, are all theorems:
\begin{align}
&\rulename{Hrec} (\forall Q : \htriple{P}{Q}{R} \simplies \htriple{P}{f(Q)}{R}) \simplies \htriple{P}{\lfp{f}}{R} \\
&\rulename{Brec} (\forall P : \btriple{\view}{P}{\vprime} \simplies \btriple{\view}{f(P)}{\vprime}) \simplies \btriple{\view}{\lfp{f}}{\vprime} \\
&\rulename{Frec} (\forall P : \ftriple{\view}{P}{\vprime} \simplies \ftriple{\view}{f(P)}{\vprime}) \simplies \ftriple{\view}{\lfp{f}}{\vprime} \\
&\rulename{Vrec} (\forall C : \vtriple{\view}{C}{\vprime} \simplies \vtriple{\view}{f(C)}{\vprime}) \simplies \vtriple{\view}{\lfp{f}}{\vprime}
\end{align}

\subsection*{Operational rules}
The operational rules for recursion rely on the fact that the least fixpoint of a function is a fixpoint, i.e. $\lfp{f} \aeq f(\lfp{f})$, and some additionally use the action \askip{} to unroll recursion. The rules of the abstract, description and command calculi look similar and all of them have straightforward proofs. Assume that $f$ is a monotone function on commands in the following rules:
\begin{align}
&\rulename{PCrec} \plotkin{f(\lfp{f})}{\s}{C}{\sprime} \simplies \plotkin{\lfp{f}}{\s}{C}{\sprime} \\
&\tag{PCrec$^\prime$} \plotkin{\lfp{f}}{\s}{f(\lfp{f})}{\s} \\
&\rulename{MCrec} \milner{f(\lfp{f})}{C}{C^\prime} \simplies \milner{(\lfp{f})}{C}{C^\prime} \\
&\tag{MCrec$^\prime$} \milner{(\lfp{f})}{\askip}{f(\lfp{f})} \\
&\rulename{KCrec} \kahn{f(\lfp{f})}{\s}{\sprime} \simplies \kahn{\lfp{f}}{\s}{\sprime}
\end{align}

\subsection*{Further remarks}
The Knaster-Tarski characterization of the least fixpoint and the Kleene star laws imply that iteration is a special case of recursion:
\begin{lemma}
$\kleene{P} \aeq \lfp{(\lambda x \,.\, \askip \aor (P \asemicolon x))}$
\end{lemma}
The monotonicity of the function $(\lambda x \,.\, \askip \aor (P \asemicolon x))$ follows from the mo\-no\-to\-ni\-ci\-ty of (\aor) and (\asemicolontext). It is possible to derive the rules for iteration by using the ones for recursion.

Users of the framework can also apply the Kleene fixpoint theorem, which provides an alternative characterization of the least fixpoint of some functions. It uses the auxiliary notions of directed sets (of languages) and Scott-continuity:
\BEGINMYDEF
$\mathit{directed}(X) \defn X \neq \emptyset \sand (\forall P \in X, Q \in X : \exists R \in X : P \aleq R \sand Q \aleq R)$ \\
$\mathit{Scott\mhyphen continuous}(f) \defn \forall X : \mathit{directed}(X) \simplies f(\alub X) \aeq \alub\{f(P) \mid P \in X\}$
\ENDMYDEF
The theorem says that if $f$ is Scott-continuous (and therefore monotone), then:
\BEGINMYDEF
$\lfp{f} = \alub\{f^n(\abottom) \mid n \in \mathbb{N}\}$
\ENDMYDEF
The composition of Scott-continuous functions is again Scott-continuous, and it is simple to prove the Scott-continuity of (\asemicolontext), (\aor), (\astar) and the Kleene star.

\end{document}